\documentclass[conference]{IEEEtran}
\IEEEoverridecommandlockouts

\usepackage{cite}
\usepackage{amsmath,amssymb,amsfonts}
\usepackage{algorithmic}
\usepackage{graphicx}
\usepackage{textcomp}
\usepackage{hyperref}
\usepackage{lipsum}
\usepackage{orcidlink}
\usepackage{makecell}
\usepackage{dblfloatfix}

\usepackage{multirow}
\usepackage{array}
\usepackage{booktabs}
\usepackage{xcolor}
\usepackage{pifont}
\usepackage{subcaption}

\usepackage{eso-pic}

\newcommand{\cmark}{\ding{51}}%
\newcommand{\xmark}{\ding{55}}%

\def\BibTeX{{\rm B\kern-.05em{\sc i\kern-.025em b}\kern-.08em
    T\kern-.1667em\lower.7ex\hbox{E}\kern-.125emX}}

\newcommand{\IEEEcopyrightnotice}{
\color{gray}
\textcopyright~2026 IEEE. Personal use of this material is permitted. Permission from IEEE must be obtained for all other uses, in any current or future media, including reprinting/republishing this material for advertising or promotional purposes, creating new collective works, for resale or redistribution to servers or lists, or reuse of any copyrighted component of this work in other works.
\smallskip\par
This paper has been accepted for publication in the Proceedings of the 2026 IEEE Radar Conference (RadarConf26).
}

\begin{document}

\title{Experimental Characterization of a Multifunction X-Band AESA Radar Demonstrator
\thanks{This work was supported by the Italian Ministry of Defense under the National Military Research Plan (PNRM) 2019, Contract No.~20539 dated December 17, 2019, within the project ``SAMBA-X - Multirole low-cost X-Band AESA Seeker for naval applications.''}
}
\author{
\IEEEauthorblockN{
Francesco Mancuso\,\orcidlink{0000-0003-4174-7816}\IEEEauthorrefmark{1},
Giulio Meucci\,\orcidlink{0000-0002-5472-2186}\IEEEauthorrefmark{1},
Matteo Pardi\,\orcidlink{0009-0007-0615-6291}\IEEEauthorrefmark{1},
Giulio Giovannetti\,\orcidlink{0000-0002-7188-6326}\IEEEauthorrefmark{2},
Alberto Lupidi\,\orcidlink{0000-0003-3964-8181}\IEEEauthorrefmark{1},
}
\vspace{3pt}
\IEEEauthorblockA{\IEEEauthorrefmark{1}Radar and Surveillance Systems (RaSS) National Laboratory - CNIT, Pisa, Italy}
\IEEEauthorblockA{\IEEEauthorrefmark{2}ELDES S.r.l., Scandicci, Italy}\\
\href{mailto:francesco.mancuso@cnit.it}{\texttt{francesco.mancuso@cnit.it}}
}

\AddToShipoutPictureFG*{
  \AtPageLowerLeft{
    \put(\LenToUnit{(\paperwidth-\textwidth)/2 - 0.05in},\LenToUnit{0.30in}){
      \parbox[b]{\textwidth}{
        \footnotesize\IEEEcopyrightnotice
      }
    }
  }
}

\maketitle

\begin{abstract}
Modern naval surveillance demands multifunction radar systems capable of operating in cluttered and contested environments. This paper presents the experimental characterization of a compact, X-band Active Electronically Scanned Array (AESA) radar demonstrator. The system was evaluated in a realistic coastal field environment at Naval Support and Experimentation Centre (CSSN) and, specifically, its specialized institute, the G. Vallauri Institute, which has historical expertise in testing and evaluating the performance of operational sensors as well as those under development, using real maritime targets and an active noise jammer. The trials assessed three core functions: direction-of-arrival (DoA) estimation, adaptive jammer suppression using MVDR beamforming, and high-resolution Inverse Synthetic Aperture Radar (ISAR) imaging. The results confirm that the demonstrator successfully detects and localizes targets, effectively suppresses high-power interference, and generates clear ISAR images of non-cooperative vessels. These findings validate the multifunction performance of the AESA demonstrator, confirming its suitability for advanced naval surveillance applications.
\end{abstract}


\section{Introduction}
Modern naval operations take place in complex environments where surveillance systems must cope with maritime traffic, sea clutter, and advanced threats such as low-observable targets, unmanned vehicles, and Electronic Countermeasures (ECM) \cite{Skolnik2008,Richards2010}. Achieving reliable situational awareness in these conditions is essential for naval platforms. Conventional radars with mechanically rotating antennas are limited by slow scan speeds and reduced agility, making them less effective against fast or multiple targets.

Active Electronically Scanned Array (AESA) technology overcomes these limitations by enabling electronic beam steering and the simultaneous formation of multiple beams, allowing a single radar to perform tasks such as search, tracking, and high-resolution imaging \cite{Mailloux2017}. The array architecture also supports adaptive beamforming for jammer suppression~\cite{Ward1994STAPReport,VanTrees2002}, while its modular structure enhances system reliability.

Although the advantages of AESA technology are well established, its real-world performance depends on factors that cannot be fully replicated in simulation, including clutter, multipath, and unpredictable interference. Field trials in representative scenarios are therefore essential to assess operational performance \cite{Haykin1985,Melvin2004}.


This paper presents the experimental characterization of a multifunction X-band AESA radar demonstrator tested in a coastal environment at the G. Vallauri Institute in Livorno, Italy. The system was evaluated against both cooperative and non-cooperative maritime targets and subjected to a high-power broadband noise jammer. The objective is to demonstrate the radar’s measured capabilities under realistic conditions. The main contributions of this work are the experimental characterization of a complete multifunction processing chain from Direction-of-Arrival (DoA) estimation to adaptive beamforming in a coastal environment; the demonstration of high-resolution Inverse Synthetic Aperture Radar (ISAR) imaging on non-cooperative maritime targets for classification; and the quantitative assessment of adaptive nulling performance confirming the radar’s effectiveness in the presence of strong electronic countermeasures. 

The remainder of the paper is organized as follows: Section~\ref{background} reviews related work, Section~\ref{methods} describes the processing algorithms, Section~\ref{setup} details the experimental setup, Section~V presents the results, and Section~\ref{conclusion} concludes the paper.

\section{Background and Related Work}\label{background}
The problem of angular estimation and radar imaging in AESA X-band systems has been extensively studied. Classical monopulse techniques guarantee high accuracy in single-target scenarios, but their performance degrades in the presence of multiple sources or strong jammers. To overcome these limitations, digital beamforming (DBF) architectures have been introduced, exploiting array processing to improve both detection and interference rejection. 

DBF can be implemented using \textit{data-independent} beamformers, based on predefined weights and array geometry, or \textit{adaptive (data-dependent)} schemes that estimate interference statistics and optimize the weight vector in real time. Among the latter, the Minimum Variance Distortionless Response (MVDR) beamformer is widely adopted, as it can place nulls in jammer directions while maintaining gain towards the target of interest~\cite{VanTrees2002,Ward1994STAPReport}. Spectral-based DoA estimators such as MUSIC further extend DBF capabilities, enabling the resolution of multiple closely spaced sources that conventional beamformers cannot distinguish~\cite{schmidt1986music}. These methods, however, require higher computational effort and precise covariance estimation. 

Architecturally, DBF can be applied at element level or at the level of \textit{sub-arrays}. While element-level processing provides maximum flexibility, sub-array solutions reduce hardware and latency requirements, which is attractive for real-time AESA applications. Comparative analyses confirm that equivalent beampatterns can be achieved if calibration is properly managed~\cite{7959180}, though increased susceptibility to grating-lobe effects remains a drawback that must be controlled through appropriate design.

Beyond angular estimation, imaging modes are increasingly relevant for target classification and situational awareness. Synthetic Aperture Radar (SAR) methods allow AESAs to form high-resolution reflectivity maps, also in forward-looking bistatic or highly squinted architectures~\cite{Chen2016MissileSAR,Chen2015ChirpScaling}; ISAR exploits target motion to achieve similar results. A critical issue in ISAR is motion compensation, since target kinematics are generally unknown~\cite{805442}. Autofocus techniques such as the Image Contrast Based Algorithm (ICBA) have become standard solutions, maximizing image sharpness through polynomial motion models~\cite{chen2014inverse,Martorella2005Contrast}.

Overall, prior research demonstrates that adaptive DBF methods (e.g., MVDR, MUSIC) combined with SAR/ISAR imaging and autofocus algorithms form the technological foundation of next-generation AESA radars. Yet, their practical implementation in real-time systems operating in complex electromagnetic environments remains a critical step toward operational maturity~\cite{937465}. This paper addresses this gap by presenting the experimental characterization of an X-band AESA demonstrator, showing that the adopted signal processing chain achieves robust and repeatable performance under realistic conditions.

\section{Methods and Algorithms}\label{methods}

The processing chain adopted in the experimental campaign includes five main steps following data acquisition:
\begin{enumerate}
    \item generation of range–Doppler (RD) matrices;
    \item digital beamforming;
    \item target detection;
    \item DoA estimation;
    \item radar imaging.
\end{enumerate}
Before beamforming, a dedicated jammer detection stage is applied. If a jammer is detected, its presence is used to trigger an adaptive cancellation process. Otherwise, standard beamforming using a fixed steering vector is employed. This chain reflects the functional sequence expected in an operational radar system.

Since automatic jammer detection was not the focus of this campaign (where the main emphasis was on beamforming, DoA estimation, and imaging), the jammer detection step was carried out manually. Specifically, the presence or absence of jamming interference was known a priori for each experiment, allowing direct selection between adaptive and conventional beamforming modes.

The entire processing chain was implemented in MATLAB and executed offline on the recorded raw data, providing flexibility for testing different operational conditions and processing configurations.

\subsection{Direction-of-Arrival Estimation}

Angular localization of both targets and interference sources was performed using the MUSIC algorithm \cite{schmidt1986music}. For each dwell, the spatial covariance matrix of the array data was computed from a training subset extracted from the corresponding RD matrix. For target DoA estimation, the training subset was selected from a region around the detected target, excluding the clutter-dominated area. MUSIC provided high-resolution angular estimates by exploiting the orthogonality between signal and noise subspaces, enabling reliable localization of both maritime targets and the broadband jammer.

\subsection{Adaptive Jammer Suppression}

To mitigate jammer interference, adaptive nulling was implemented through MVDR beamforming. For the sample covariance matrix estimation localization, the training subset was taken from a region dominated by interference energy. For each look direction, the beamforming weights were computed from the sample covariance matrix so as to preserve unit gain toward the desired signal direction while minimizing the total array output power. This adaptive response effectively produced deep nulls in the jammer direction, significantly reducing interference energy at the array output. The beamforming weights were always normalized to have unit norm in order to preserve the noise floor level. For reference, non-adaptive beamscans were also generated using fixed steering vectors, providing a baseline for assessing the benefit of the adaptive approach.

\subsection{Radar Imaging}

High-resolution imaging of detected vessels was obtained using ISAR techniques \cite{chen2014inverse}. The ISAR processing exploited the relative motion of maritime targets to synthesize a two-dimensional image from a sequence of range–Doppler maps. This allowed detailed characterization of target structure and motion, complementing the detection and localization results.

\section{Experimental Setup}\label{setup}

\begin{figure}[t!]
    \centering
    \includegraphics[width=0.8\linewidth]{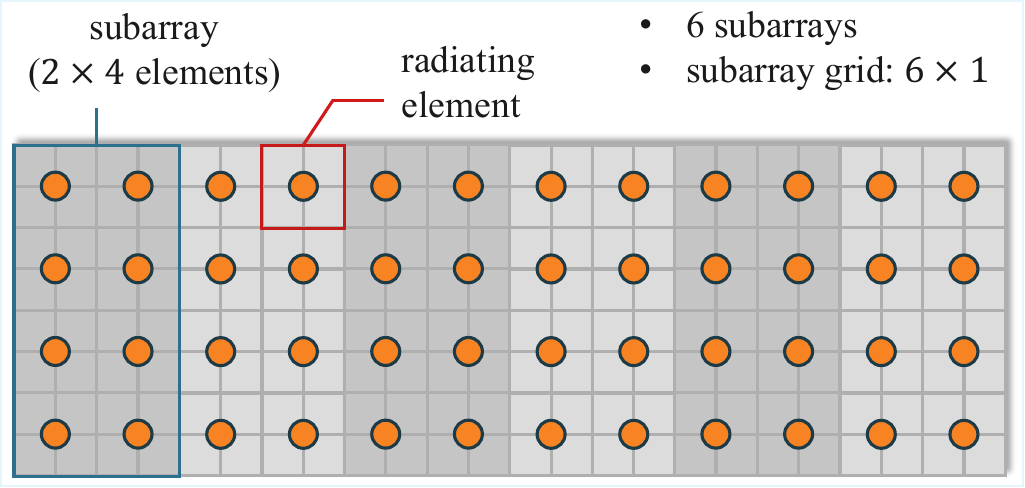}
    \caption{Array schematic of the radar demostrator.}
    \label{fig:array-scheme}
\end{figure}

This section describes the radar demonstrator and the experimental campaign, with emphasis on the methodologies adopted, the operational conditions, and the objectives of each test. Fig.~\ref{fig:array-scheme} shows the array configuration of the radar demonstrator. The system is an X-band AESA radar composed of 48 radiating elements arranged in a $12 \times 4$ rectangular grid (azimuth $\times$ elevation), with a spacing of $\lambda/2$ between elements. The array is partitioned into six subarrays, each consisting of $2 \times 4$ elements and equipped with its own receiver. This architecture provides six independent receiving channels, enabling flexible digital beamforming and adaptive interference mitigation experiments.

\begin{figure}[ht]
    \centering
    \begin{subfigure}{\linewidth}
        \hspace{5pt}
        \includegraphics[width=\linewidth]{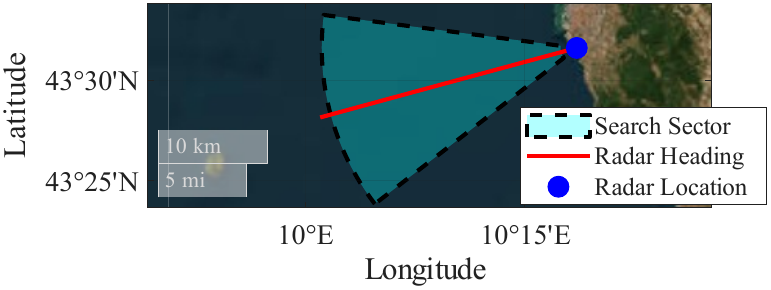}
        \subcaption{Geographical view showing the radar position, heading, and 45° search sector.}
    \end{subfigure}\\\vspace{5pt}
    \begin{subfigure}{\linewidth}
        \includegraphics[width=\linewidth]{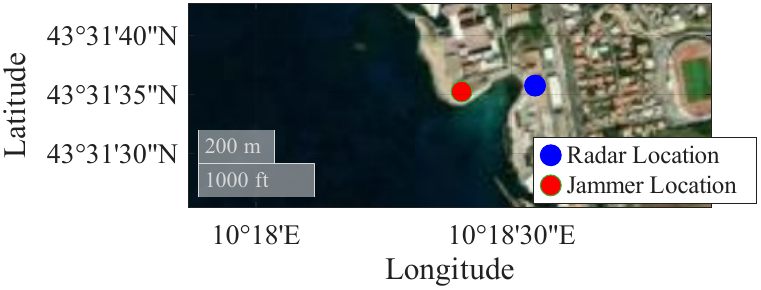}
        \subcaption{Detail of the radar and jammer locations.}
    \end{subfigure}
    \caption{Operational scenario of the experimental campaign.}
    \label{fig:map}
\end{figure}

The radar was installed on a fixed coastal platform, oriented with a heading of $252^\circ$ with respect to north. The sensor covered an azimuth sector of approximately $45^\circ$ and a range interval between 1.5 km and 23.5 km. The overall geometry of the radar site and the jammer position within the test area are shown in Fig.~\ref{fig:map}. The jammer used for nulling experiments was located about $203$ m from the radar, at an azimuth of $21.4^\circ$  relative to the platform reference. This short baseline created a highly stressing scenario, with Jammer-to-Noise Ratios (JNR) of approximately $50$ dB, suitable for assessing the robustness of adaptive interference suppression algorithms. Automatic Identification System (AIS) data were collected in parallel and used as external ground truth to validate detection and DoA estimation results.

\begin{table}[!b]
\centering
\caption{Overview of the conducted tests with targets, jammer conditions and algorithms used.}
\renewcommand{\arraystretch}{1.5}
\setlength{\tabcolsep}{4pt} 
\small 

\begin{tabular}{@{}c l c c r@{}}
\toprule
\textbf{ID} &
\makecell[l]{\textbf{Title}} &
\textbf{Target} &
\textbf{Jammer} &
\makecell{\textbf{Algorithm}} \\
\midrule
\textbf{T1} & \makecell[l]{Angular detection\\on single target}                             & \cmark & \xmark & MUSIC \\
\textbf{T2} & \makecell[l]{High-power jammer\\cancellation}                                & \cmark & \xmark & MVDR \\
\textbf{T3} & \makecell[l]{Target detection with\\high-power jammer}                       & \cmark & \cmark & \makecell[r]{MVDR,\\MUSIC} \\
\textbf{T4} & \makecell[l]{Imaging single target\\with multiple scatterers}                & \cmark & \xmark & ISAR \\
\bottomrule
\end{tabular}
\label{tab:tests}
\end{table}

The measurement activities focused on several key capabilities of the radar system: (i) DoA estimation of targets, (ii) high-power jammer cancellation, (iii) target detection with high-power jammer, and (iv) radar imaging quality. Tab.~\ref{tab:tests} summarizes the test cases.

\subsection*{T1 -- Angular detection on single target}

This test assessed the system’s capability to estimate the DoA of both cooperative and non-cooperative targets. No jammer was present during these experiments. For each measurement, a detected maritime target was selected and its DoA was estimated using the MUSIC algorithm. The radar-derived bearing was then compared against the corresponding AIS-reported position, used as ground truth. The procedure was repeated for multiple targets and measurement sessions to verify the consistency and accuracy of the DoA estimation. The expected outcome was a small angular deviation between the radar-derived and AIS-derived DoA measurements.

\subsection*{T2 -- High-power jammer cancellation}

This test evaluated the system’s capability to suppress a high-power barrage jammer while maintaining gain in the desired direction. The data collected in each dwell were stored as multichannel RD datacubes and processed using two approaches: (i) conventional non-adaptive beamforming and (ii) adaptive MVDR beamforming. For each pointing direction, the MVDR algorithm computed adaptive weight vectors that shaped the antenna response to minimize interference while preserving the mainlobe gain. The expected outcome was a strong jammer suppression in the MVDR-processed data compared to the non-adaptive case.

\begin{table*}[!b]
\centering
\caption{Angular errors across all sessions. Asterisks denote vessels oriented nearly parallel to the radar line-of-sight. The \emph{Within Target?} column shows whether the error falls within the target span, consistent with expected angular glint effects.}
\renewcommand{\arraystretch}{1.1}
\setlength{\tabcolsep}{12pt}
\begin{tabular}{@{}lcccc|ccc@{}}
\toprule
\textbf{Session} & 
\makecell{\textbf{Detection}\\\textbf{Range (km)}} &
\makecell{\textbf{Ship}\\\textbf{Size (m)}} &
\makecell{\textbf{Ship}\\\textbf{Heading (°)}} &
\makecell{\textbf{Projected}\\\textbf{Size (m)}} &
\makecell{\textbf{Target Angular}\\\textbf{Span (°)}} &
\makecell{\textbf{Angular}\\\textbf{Error (°)}} &
\makecell{\textbf{Within}\\\textbf{Target?}} \\ 
\midrule
Eurocargo Genova \cite{eurocargoGenova}          & 10.15 & 200 & 9.1   & 178 & 1.0    & 0.1  & \cmark \\
Mega Express \cite{megaExpress}              & 7.68   & 176 & 35.3  & 105 & 0.8 & 0.1  & \cmark \\
Mega Express               & 8.22      & 25* & 221.2     & 25*            & 0.2              & 0.9  & \xmark \\
Rossetti \cite{marina_raffaele_rossetti}                   & 15.78 & 8*  & 254.0   & 8*             & 0.03 & 0.2  & \xmark \\
Stelio Montomoli \cite{stelio_montomoli}           & 6.82      & 93  & 36.0     & 55             & 0.5              & 0.2  & \cmark \\
Zeus Palace \cite{zeusPalace}                & 13.42      & 212 & 200.8     &  165            & 0.7              & 0.4  & \cmark \\
Mega Smeralda \cite{megaSmeralda}              & 8.01      & 171 & 37.7     & 96             & 0.7              & 0.2  & \cmark \\
Epaminondas \cite{epaminondas}                & 12.57      & 43* & 32.5     & 43*             & 0.2              & 0.2  & \cmark \\
\bottomrule
\end{tabular}
\label{tab:session_errors}
\end{table*}

\subsection*{T3 -- Target detection with high-power jammer}

This test investigated the radar’s ability to detect a target and estimate its DoA in the presence of an active, high-power broadband noise jammer. Prior to Range–Doppler map inspection, an MVDR beamforming stage was applied to the multichannel RD datacube to attenuate the jammer and enhance target visibility. After identifying the target on the MVDR-enhanced RD map, the MUSIC algorithm was applied to the original RD data to estimate the DoA. The estimated bearing was then compared with the AIS-derived reference for the same target. The expected results were effective jammer suppression, successful target detection, and a small angular deviation between radar estimates and AIS measurements.

\subsection*{T4 -- Imaging single target with multiple scatterers}

This test examined the system’s capability to generate ISAR images of non-cooperative targets and to resolve multiple scattering centers. No jammer was active during this experiment, and a conventional beamforming with a fixed steering vector was used. The ISAR processing chain consisted of three steps: (1) generating the Range–Doppler map and isolating the target return, (2) applying ISAR processing with motion compensation and Doppler integration, and (3) forming a focused two-dimensional ISAR image. The expected result was a high-resolution image revealing multiple scattering centers that characterize the target’s geometry.

\section{Results and Discussion}\label{results}
The following sections present the experimental results obtained for each of the four test cases summarized in Tab.~\ref{tab:tests}. Each test investigates a specific functional aspect of the radar and provides evidence of performance under realistic operating conditions. The analysis combines both quantitative and qualitative results: quantitative metrics are used to assess measurable parameters such as angular accuracy and jammer rejection, while qualitative evaluations illustrate detection capability and image quality through representative Range–Doppler and ISAR products. Together, these results offer a comprehensive characterization of the radar demonstrator, covering DoA estimation, adaptive interference suppression, target detection in jamming conditions, and high-resolution imaging.

\subsection*{T1 -- Angular detection on single target}
\begin{figure}[ht]
    \centering
    \includegraphics[width=\linewidth]{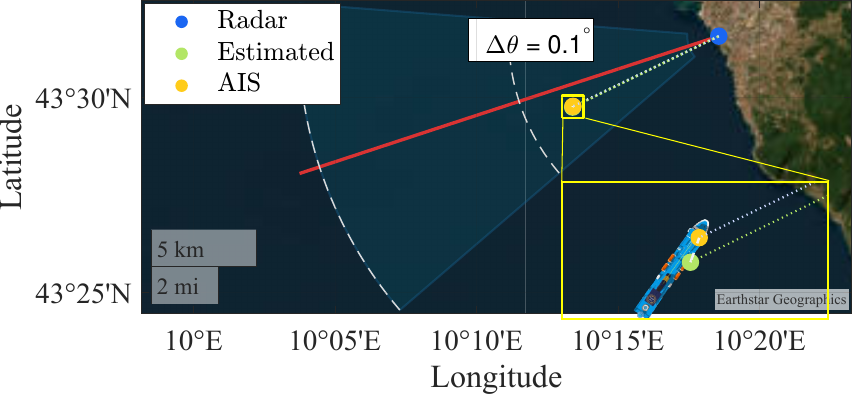}
    \caption{Example of angular detection on a maritime target. The ship is adequately scaled and georeferenced.}
    \label{fig:megaespressDOA}
\end{figure}

Fig.~\ref{fig:megaespressDOA} shows an example of angular detection performed on a non-cooperative target. The radar-derived DoA estimate is closely aligned with the AIS-reported position, confirming the system’s accuracy under interference-free conditions.

Quantitative results for all test sessions are summarized in Tab.~\ref{tab:session_errors}. The average angular deviation across all trials was $0.28^{\circ}$. All values reported under \emph{Projected Size} correspond to the apparent cross-range extent of each vessel as seen by the radar, i.e., the portion of the ship’s physical length obtained by projecting its heading vector onto the direction perpendicular to the radar line-of-sight (cross-range axis). Values marked with an asterisk refer instead to the vessel’s actual beam (true width), which occurs when the ship’s heading is nearly parallel to the radar line-of-sight. In this configuration the vessel presents its narrow side to the radar, so the projected length collapses and the apparent cross-range extent becomes close to the beam. Overall, the results demonstrate a stable sub-degree angular precision consistent with the angular extent of the target as observed from the radar, with slightly larger deviations observed for near-parallel headings, where the effective radar cross-section and angular span are reduced. 

\subsection*{T2 -- High-power jammer cancellation}
\begin{figure}[ht]
\centering
\includegraphics[width=\linewidth]{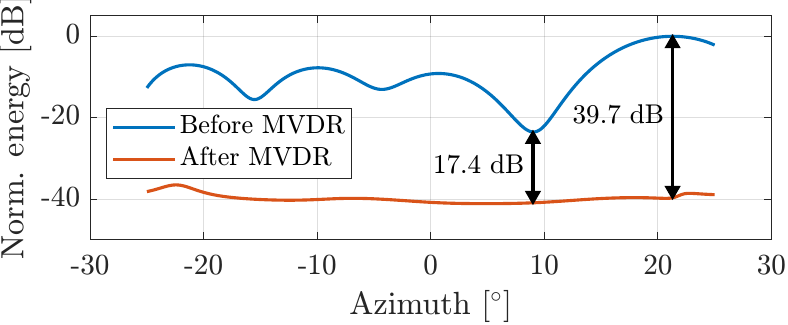}
\caption{Normalized received energy versus steering angle before and after MVDR.}
\label{fig:JammerCancVSAngle}
\end{figure}

The results summarized in Tab. \ref{tab:jammer_rejection} confirm the effectiveness of the adaptive MVDR algorithm in suppressing the broadband noise jammer across the different steering configurations. An average rejection level above 30~dB was achieved, with peaks almost reaching 40~dB. The lowest rejection value was caused by an inherent null or minimum in the jammer’s direction when using the conventional non-adaptive beamformer. This can be visually verified by analysing the energy measured before applying the MVDR algorithm, as shown in Fig.~\ref{fig:JammerCancVSAngle}.
\begin{table}[ht]
\centering
\caption{Measured rejection levels at different steering angles.}
\renewcommand{\arraystretch}{1.2}
\setlength{\tabcolsep}{5pt}
\begin{tabular}{@{}cccccc@{}}
\toprule
\textbf{-20°} & \textbf{-10°} & \textbf{0°} & \textbf{10°} & \textbf{20°} & \makecell{\textbf{Average}\\\textbf{Rejection}} \\ 
\midrule
30.9~dB & 32.4~dB & 31.6~dB & 18.8~dB & 39.5~dB & 30.6~dB \\
\bottomrule
\end{tabular}
\label{tab:jammer_rejection}
\end{table}

\begin{figure}[ht]
\centering
\includegraphics[width=\linewidth]{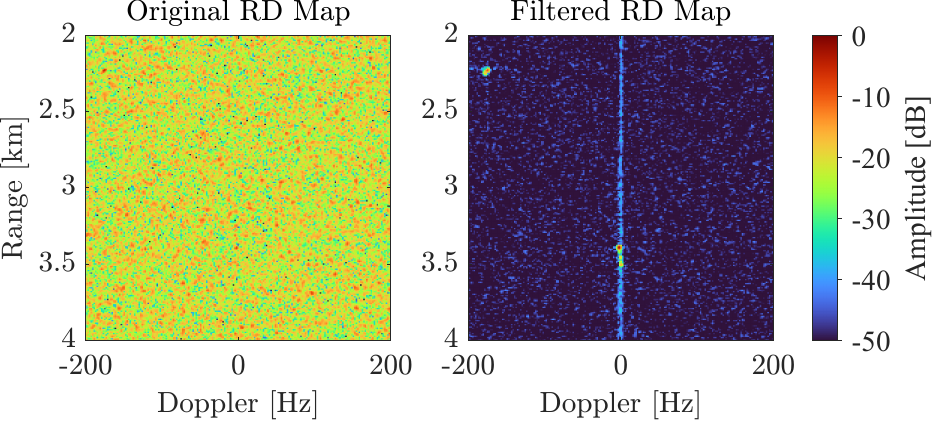}
\caption{Range–Doppler maps before and after adaptive jammer cancellation.}
\label{fig:jammerCanc}
\end{figure}

Fig. \ref{fig:jammerCanc} illustrates a representative example for the $0^\circ$ (broadside) steering direction. The two Range–Doppler maps, obtained respectively with conventional beamforming and with MVDR, are normalized with respect to the overall maximum value computed across both maps, in order to preserve a consistent visual comparison. By inspection, the background noise floor is reduced by approximately 30~dB after the adaptive cancellation, revealing two distinct targets that were not visible before: one stationary at about 3.5~km, and a second one moving away from the radar at roughly 2.2~km. These results highlight the ability of the MVDR algorithm to strongly attenuate the jammer contribution while preserving the visibility of targets in the scene.

Fig.~\ref{fig:JammerCancVSAngle} provides a broader view of the suppression performance as a function of the array steering angle. The normalized energy of the received signal is plotted for both the conventional non-adaptive beamformer (in blue) and the MVDR (in red) beamforming approaches. The results demonstrate that the jammer contribution, dominant under conventional beamforming, is consistently attenuated by about 30~dB across most directions, confirming the robustness of the adaptive cancellation.

\subsection*{T3 -- Target detection with high-power jammer}
\begin{figure}[h!]
\centering
\includegraphics[width=\linewidth]{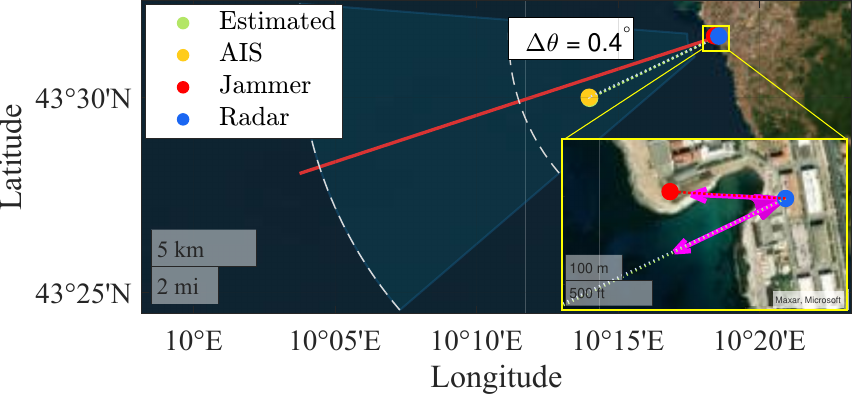}
\caption{Target detection in the presence of an active jammer.}
\label{fig:DetectionWithJammer}
\end{figure}

Fig.~\ref{fig:DetectionWithJammer} illustrates the detection of a cooperative vessel in the presence of an active jammer. An intermediate MVDR filtering stage was applied to suppress the jammer and highlight the target in the Range–Doppler map, similarly as demonstrated for test T2 in Fig.~\ref{fig:jammerCanc}. The MUSIC algorithm was then applied to a patch around the detection using the original (non-filtered) data in order to preserve the angular integrity of the target echo. The magenta overlay represents the MUSIC spatial spectrum, where arrows mark the two most dominant peaks corresponding to the estimated directions of arrival. The radar-estimated DoA was compared with the AIS measurement. These findings demonstrate that the combined processing chain is effective in maintaining reliable target detection despite high-power interference.

\subsection*{T4 -- Imaging single target with multiple scatterers}
To evaluate the imaging performance of the radar system, two maritime targets mentioned above were selected and observed under realistic operating conditions. For each vessel, an optical reference image is presented alongside the corresponding ISAR reconstruction to allow a direct qualitative comparison. 

\begin{figure}[h!]
    \centering
    \hspace{12pt}
    \begin{subfigure}{0.45\linewidth}
        \centering
        \includegraphics[width=\linewidth]{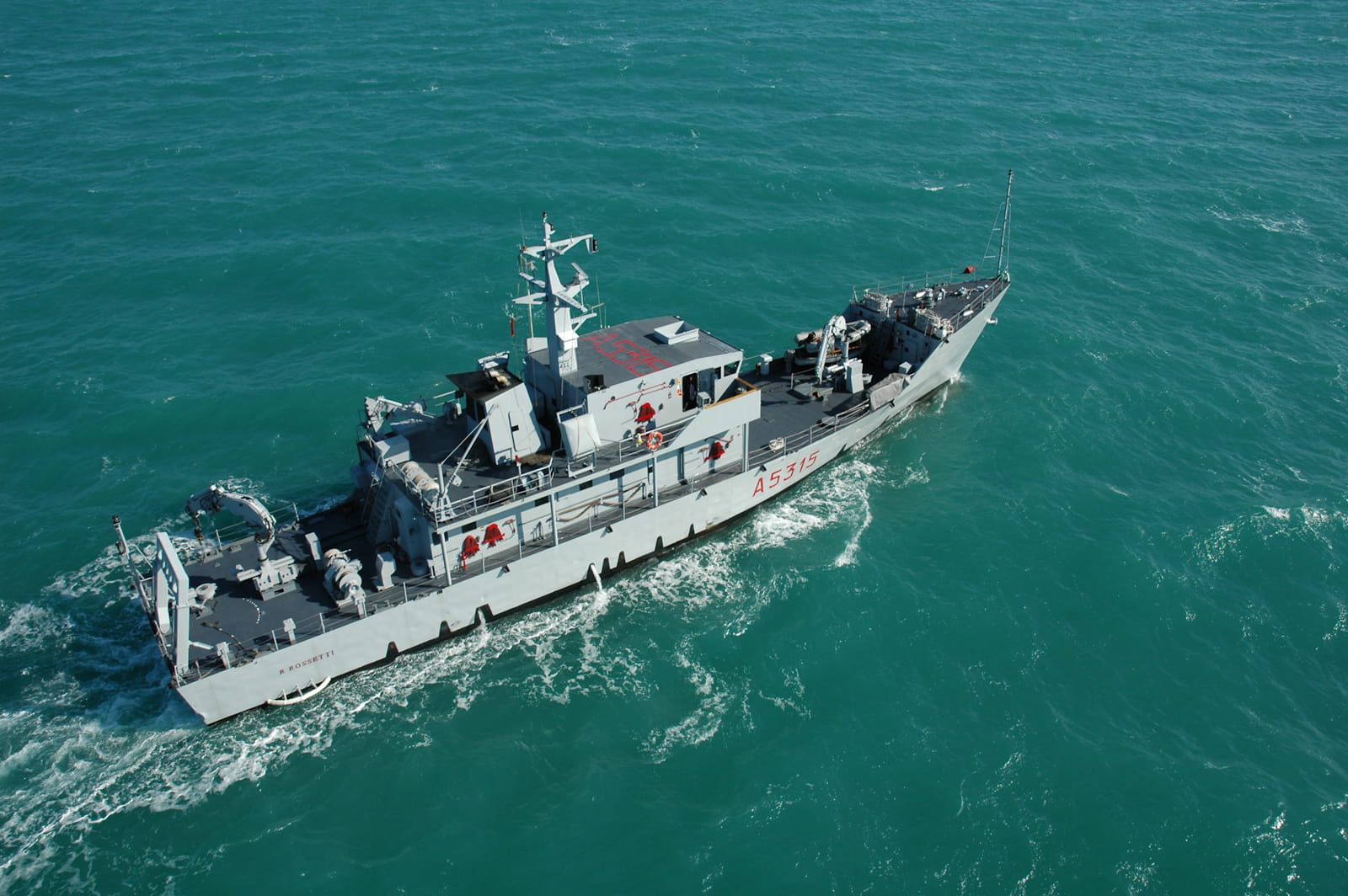}
        \subcaption{Rossetti – optical image.}
    \end{subfigure}
    \hspace{3pt}
    \begin{subfigure}{0.45\linewidth}
        \centering
        \includegraphics[width=\linewidth]{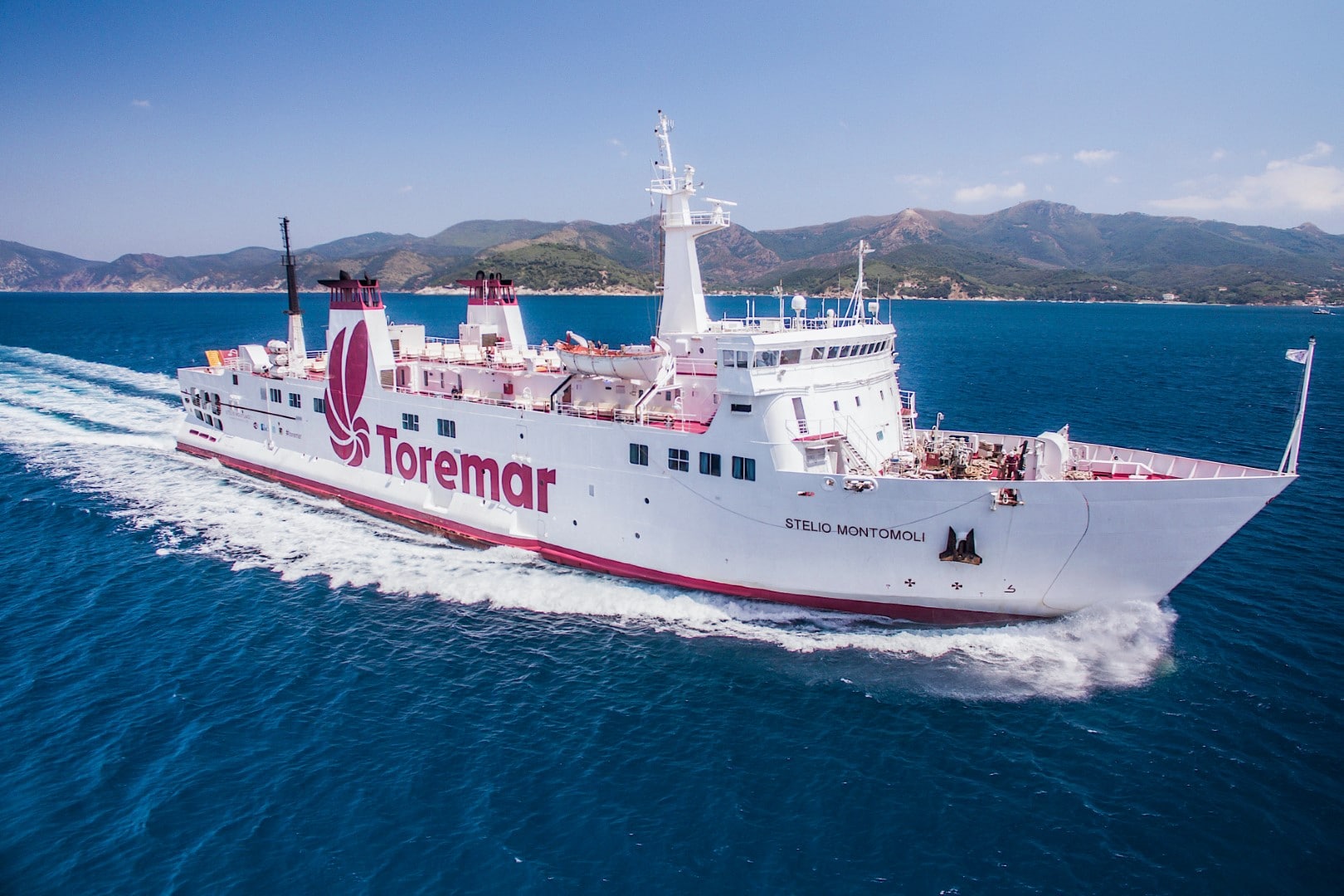}
        \subcaption{Montomoli – optical image.}
    \end{subfigure}

    \vspace{10pt}

    \begin{subfigure}{0.48\linewidth}
        \hspace{-2.5em}
        \raisebox{0.5em}{%
            \includegraphics[height=3cm]{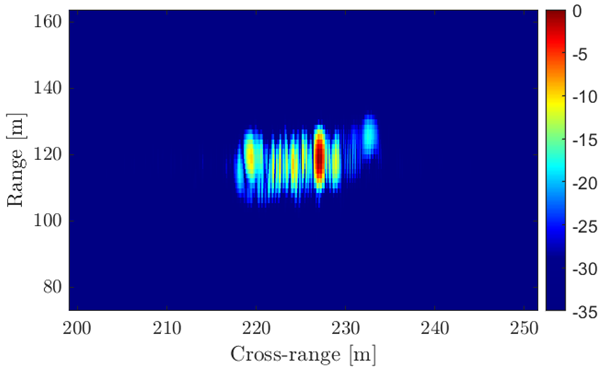}
        }
        \subcaption{Rossetti – ISAR image.}
    \end{subfigure}
    \hspace{12pt}
    \begin{subfigure}{0.44\linewidth}
        \centering
        \includegraphics[height=3.3cm]{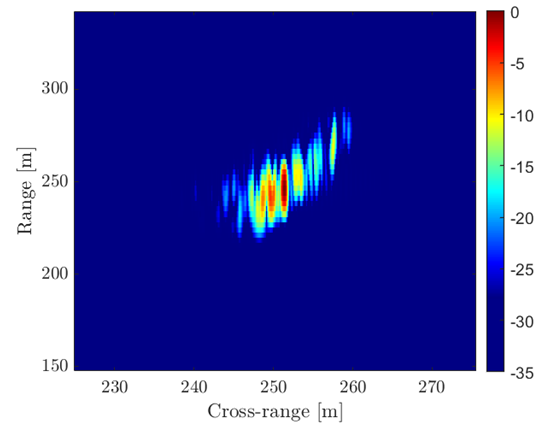}
        \subcaption{Montomoli – ISAR image.}
    \end{subfigure}

    \caption{
        Optical and ISAR imaging of the naval vessel Raffaele Rossetti
        (Length $\times$ beam = $45\times8$ m), and the passenger ferry Stelio Montomoli (Length $\times$ beam = $93\times16$ m). Colorbars are normalized to the maximum value of each ISAR image and are expressed in dB.
    }
    \label{fig:isar_2x2}
\end{figure}

The naval ship (Fig. \ref{fig:isar_2x2}(a)) produces an ISAR image (Fig. \ref{fig:isar_2x2}(c)) characterized by a compact scattering distribution, with localized reflections that can be attributed to the superstructure and mast-like elements. Although the radar-derived spatial extent of the vessel is smaller than its true dimensions, the preserved scattering centers highlight the radar’s capability to capture the main structural features. The passenger ferry (Fig. \ref{fig:isar_2x2}(b)), in contrast, exhibits a broader set of returns with high-intensity scattering centers distributed along the length of the hull.


The ISAR image of the passenger ferry (Fig. \ref{fig:isar_2x2}(d)) reflects the elongated geometry and larger radar cross-section of the vessel, with multiple scattering centers distributed along the hull and superstructure. However, as in the previous case, the reconstructed extent of the target appears systematically smaller than its actual dimensions. This underestimation can be ascribed to a combination of physical and processing factors. Self-occlusion and scintillation can cause parts of the structure to disappear or be shadowed, reducing the apparent scattering area. Residual motion-compensation errors also contribute: in ISAR, the cross-range scaling depends on the estimated rotational rate $\Omega$, so any overestimation, caused by mismatched or non-uniform target motion, compresses the cross-range axis. Moreover, the intrinsic 2D ISAR ambiguity adds distortion, since the image plane is tied to the instantaneous rotation vector, whose exact orientation is uncertain. Despite these factors, the reconstructions consistently capture the dominant scattering centers, providing a meaningful structural representation of the target.


\section{Conclusions}\label{conclusion}
This paper presented the experimental characterization of a multifunction X-band AESA radar demonstrator, validated through coastal field trials at G. Vallauri Institute in Livorno, Italy. Four test cases were conducted to assess DoA estimation, adaptive jammer suppression, target detection under interference, and high-resolution ISAR imaging.

The results confirmed sub-degree angular accuracy consistent with the angular extent of the target as observed from the radar, and an average jammer rejection above 30~dB, demonstrating the effectiveness of the MVDR adaptive beamformer. In the presence of high-power broadband interference, the radar maintained reliable detection performance, while interference-free trials enabled the generation of clear ISAR images of non-cooperative vessels, revealing multiple scattering centers consistent with the targets’ structure, even if the cross-range dimension suffers from scaling inaccuracies and projection ambiguities inherent to ISAR imaging.

These outcomes validate the radar’s multifunction operation under realistic conditions. Future work will focus on real-time implementation of adaptive processing, refinement of imaging modes to support target classification, and extension of tracking capabilities in dynamic maritime environments.

\section*{Acknowledgment}
The authors are deeply grateful to the Italian Navy for making available ITS R. Rossetti and for access to the G. Vallauri Institute facilities, both of which were instrumental to the success of the experimental campaign. Furthermore, the authors would like to acknowledge the Italian General Secretariat of Defence, with special thanks to the Directorate of Naval Armaments, for their outstanding leadership of the program, guaranteeing the fulfillment of all established goals.

\bibliographystyle{IEEEtran} 
\bibliography{references} 

\end{document}